# Phase Transition Mechanism of Hexagonal Graphite to Hexagonal and Cubic Diamond: Ab-initio Simulation

Ranjan Mittal[1,2], Mayanak Kumar Gupta[1] and Samrath Lal Chaplot[1,2]

[1]*Solid State Physics Division, Bhabha Atomic Research Centre, Mumbai, 400085, India*
[2]*Homi Bhabha National Institute, Anushaktinagar, Mumbai 400094, India*

We have performed ab-initio molecular dynamics simulations to elucidate the mechanism of the phase transition at high pressure from hexagonal graphite (HG) to hexagonal diamond (HD) or to cubic diamond (CD). The transition from HG to HD is found to occur swiftly in very small time of 0.2 ps, with large cooperative displacements of all the atoms. We observe that alternate layers of atoms in HG slide in opposite directions by (1/3, 1/6, 0) and (-1/3, -1/6, 0), respectively, which is about 0.7 Å along the ±[2, 1, 0] direction, while simultaneously puckering by about ±0.25 Å perpendicular to the a-b plane. The transition from HG to CD occurred with more complex cooperative displacements. In this case, six successive HG layers slide in pairs by 1/3 along [0, 1, 0], [-1, -1, 0] and [1, 0, 0], respectively along with the puckering as above. We have also performed calculations of the phonon spectrum in HG at high pressure, which reveal soft phonon modes that may facilitate the phase transition involving the sliding and puckering of the HG layers. The zero-point vibrational energy and the vibrational entropy are found to have important role in stabilizing HG up to higher pressures (>10 GPa) and temperatures than that estimated (<6 GPa) from previous enthalpy calculations.

The compression of graphite under extreme thermodynamic conditions is subject of intense scientific research [1-6]. Cubic diamond (CD) was recovered[3] after explosive shock compression of hexagonal graphite (HG). Synthetic diamond can be obtained by simultaneously applying high pressure and temperatures on graphite [1,3,7]. Hexagonal form of diamond[8], which was later called lonsdaleite[4], was identified inside fragments of the Canyon Diablo meteorite. Lonsdaleite [4] is believed to be a harder form of carbon than cubic diamond[9,10]. Static compression of hexagonal graphite at ~13-14 GPa results in transformation to hexagonal diamond (HD) [8,11] or CD [2,12] at temperatures around 1000 K.

Recent shock studies [13-16] of the transition from HG to HD or CD have created much interest in understanding the mechanism of these transitions. A variety of experiments have been performed, namely, using a laser shock [13], plate impact shock [14,15], and detonation-induced shock experiments [16], which reveal the transition above 50 GPa.

The thermodynamic behaviour of graphite as well as that of other graphitic materials, such as graphene and carbon nanotubes is of considerable scientific interest [17-28]. Theoretical works have been reported[29-33] to understand the dynamic process for synthesis of CD and HD. Various theoretical studies of the enthalpy in various phases indicated[34,35] that under compression above a few GPa, both CD and HD are more stable in comparison to HG, while CD is always more stable than HD. However, experimental measurements [8,11] [13-16] also show transformation from HG to HD.

The measurements of phonon modes in HG and CD have been reported [36-39] using Raman, infrared, neutron as well as x-ray inelastic scattering techniques. First-principles investigation of the structural, vibrational and thermodynamic properties of diamond and graphite have also been reported [40].

The unit cell of hexagonal graphite structure (space group P6$_3$/mmc) consists of 4 carbon atoms that occupy two Wycoff sites, namely, 2b {(0,0,1/4), (0,0,3/4)} and 2c {(1/3,2/3,1/4), (2/3,1/3,3/4)}. Interestingly, hexagonal diamond also crystalizes in P6$_3$/mmc space group, but the carbon atoms occupy 4f Wycoff sites (1/3,2/3,z), (2/3,1/3,z+1/2), (2/3,1/3,-z) and (1/3,2/3,-z+1/2), with z=0.0625. The cubic diamond crystallizes in Fd-3m space group. The structure of cubic diamond can also be represented in a hexagonal unit cell containing 6 atoms (see Supplementary Material [41]).

Despite numerous previous experimental as well as theoretical studies, atomic level understanding of the transformation from HG to HD/CD remains incomplete. The calculations of the energy barrier [16] do not always give the complete picture including correlated dynamics. The nucleation and growth mechanism of these phase transitions has been investigated in detail [29,32,33]. However, the transition in shock experiments occurs at a fast time scale of ~ps, and may involve a different mechanism. We have performed ab-initio molecular dynamics (AIMD) simulations to understand the mechanism of the transition from HG to HD as well as to CD. Our simulations suggest that, once the lattice is sufficiently compressed, the transition can occur swiftly through cooperative movement of all the atoms, unlike in the nucleation mechanism. The simulations thus seem to provide an alternative mechanism that may be applicable to the swift transitions observed in shock experiments. Such a concerted mechanism has been suggested earlier[42,43]. However, the present AIMD simulations provide evidence of clear atomic dynamics, the pathways and the time scales.

The ab-initio enthalpy calculations [16,34] based on the density-functional theory indicated the transition from HG to HD/CD at a rather low pressure of ~1 to 6 GPa depending on the choice of the density functional, which is much below the experimental transition pressure. The phonon spectra have important role in determining the phase stability, through the vibrational energy including the zero-point vibrational energy, and the vibrational entropy. We have calculated the equation of state and the phonon spectra at high pressures using ab-initio methods, and used these to calculate the Gibbs free energies in all the three phases to understand the role of atomic dynamics in phase transitions. The computational details are given in Supplementary Material [43]. We find that many of the phonon modes in HG occur at rather low energies than those in HD/CD. Such low-energy modes in HG contribute to increase in the vibrational entropy, thereby favoring the stability of HG over HD/CD. As a result, for example, at 2000 K, the HG is found to be more stable than HD/CD up to 14 GPa, which is close to the experimental values. We have further identified soft phonon modes in HG at high pressures which could be related to the phase transition mechanism.

**Equation of state**

The experimental data of equation of state for HG [44,45], HD [14,16], and CD [46]) are available in the literature. The calculated equation of state is compared with the experimental data in **Fig. 1(c).** The calculated volume drop of about 13 % at 49 GPa on transition from HG to HD is in agreement with the experimental data from literature [14,16]. The calculated bulk modulus values for HG and CD/HD are 38 GPa and 437 GPa, respectively, which are in good agreement with the available experimental data of 35.8 for HG [47] and 443 GPa for CD [48]. The large difference in the experimental bulk modulus values of HG and CD are very well reproduced by our calculations. Further, for HG there is a large difference in linear compressibility along the a- and c-axes, which is also in good agreement with the experimental data for HG [44,45].

**Phonon Spectra**

The calculated phonon spectra (**Fig. 1(a,b)**) show significant difference among HG, HD and CD, which mainly arises from the difference in nature of atomic bonding. The HG exhibits large anisotropy with strong $sp^2$ covalent bonding in the a-b plane, and van der Waals interactions between the graphite layers along c-axis. The covalent bonding in the a-b plane results in peaks in the phonon spectrum at about 58 meV, 78 meV and 170 meV. However, there are low energy modes below about 20 meV due to the van der Waals interactions, which are of particular interest as these modes give rise to large vibrational entropy. The calculated phonon spectrum for HG is in very good agreement with the experimental data[37] from neutron inelastic scattering.

The HD and CD involve isotropic $sp^3$ bonding. The broad peak in the phonon spectrum (**Fig. 1(b)**) of CD at around 80 meV and peaks at around 122 meV and 147 are in good agreement with the available experimental data [39]. The calculated phonon spectrum of HD is close to that of CD (**Fig. 1(b)**).

**Calculated Gibbs Free Energy in Various Polymorphs**

We have calculated the phonon spectra at several pressures in all the three phases and calculated the Gibbs free energy as a function of pressure and temperature. We find that the inclusion of vibrational energy and entropy contribution leads (**Fig. 1(c)**) to stability of cubic diamond above 10 GPa at 10 K, and at higher pressure of 14 GPa at 2000 K. It is evident that the inclusion of the phonon contribution leads to upwards shift in transition pressure, which is much closer to the experimental values, and which removes the discrepancy between the experiments and previous enthalpy calculations.

The enthalpy and free energy calculations show that CD is always more stable in comparison to HD. It may be noted that the unit cell volume of CD is slightly smaller than that of HD. For example, at 49 GPa the volumes of CD and HD per atom are 5.19 Å$^3$ and 5.20 Å$^3$, respectively; the essential difference is that while all the C-C bonds in CD are of 1.50 Å, in HD three C-C bonds are of 1.50 Å, while the fourth C-C bond along the c-axis is of 1.52 Å. The slightly smaller volume for CD may occur due to the different layer sequence along the hexagonal-axis and favourable long-range van der Waals attraction for CD.

**Phase Transition Mechanism of Hexagonal Graphite to Hexagonal and Cubic Diamond**

The AIMD simulations are performed on the 4×4×1 supercell of HG at various temperatures from 300 K to 6000 K at constant volume and temperature using NVT ensemble. The volume chosen is that of HD at 49 GPa. In the simulations at temperatures up to 5000 K, we observe fairly large shifts of atoms from their reference positions in HG, which indicate a tendency towards the transition to HD, but the transition did not complete at a time scale of 20 ps. However, at 5500 K, the calculated time dependence of fractional coordinates of the carbon atoms clearly shows (**Fig. 2**) the HG to HD transition. The transition of HG to HD at 5500 K occurs in very short times of less than 1 ps. As shown in **Fig. 2,** during the transition there is cooperative motion of all the four carbon atoms in the unit cell.

The calculated snapshots of atomic coordinates at various times are shown in **Fig. 3** (For more details see **Fig. S4** (Supplementary Material [43]). The structure of HG has two layers of carbon atoms. The time evolution of the coordinates as well as snapshots of atoms at various time steps show that all the atoms in both the layers move simultaneously. We find that there is large displacement of all the atoms (**Fig. 2**) along all the three axes (x, y and z) in first 200 fs. As shown in **Fig. 3** (For more details see **Fig. S4** Supplementary Material [43])**,** all the atoms acquire 4-fold coordination in less than about 0.5 ps. In about 0.5 ps, the atoms, which are initially in HG at {(0 0 ¼), (1/3 2/3 ¼), (0 0 ¾) and (2/3 1/3 ¾)}, shift by {(1/3, 1/6, 0.06), (1/3, 1/6, -0.06), (-1/3, -1/6, 0.06), (-1/3, -1/6, -0.06)}, respectively. It is clear that each layer of atoms in graphite slides by about 0.7 Å along ±[2, 1, 0] direction while simultaneously puckering by about ±0.25 Å perpendicular to the a-b plane. The final coordinates correspond to the standard coordinates in HD when the origin of the unit cell is appropriately shifted by (0, 1/2, -1/4). Initially in HG structure carbon atoms have three-fold co-ordination, and on transformation to HD the atoms acquire four-fold coordination in agreement with the experimental structure [4]. Our AIMD simulation clearly brings out a very swift displacive phase transition at ps time scale at high pressure and temperature akin to that found in shock experiments [15,16].

In order to simulate the transition from HG to CD, we have performed AIMD simulations on the 3×3×3 supercell of HG with NVT ensemble. The volume used is that of CD at 49 GPa. We could observe the transition at 300 K in a fairly short time of less than about 0.5 ps. The transition path involves (**Fig. 2**) cooperative movement of all the atoms. The six successive HG layers' slide in pairs by 1/3 along [1, 0, 0] and equivalent directions, namely, (0, 1/3, 0), (-1/3, -1/3, 0), (-1/3, -1/3, 0), (1/3, 0, 0), (1/3, 0, 0), and (0, 1/3, 0), respectively, while simultaneously puckering by about ±0.25 Å perpendicular to the a-b plane. We find that each layer of atoms in graphite slides by about 0.8 Å along with puckering by about ±0.25 Å.

It appears that the transition paths of HG to HD and HG to CD are quite similar, both involving swift cooperative movement of all the atoms that includes sliding of the graphite layers by about 0.7 Å with simultaneous puckering of the layers by about ±0.25 Å in a short time of less than 0.5 ps, about 0.2 ps. The direction of the sliding of HG layers is different in the two cases. We observe that in the transition from HG to HD, alternate HG layers slide in opposite directions by about 0.7 Å along ±[2, 1, 0]. The transition from HG to CD involves sliding of six successive HG layers in pairs by ~0.8 Å along [0, 1, 0], [-1, -1, 0] and [1, 0, 0], respectively. The essential difference between HD and CD is the periodicity along the hexagonal direction; it is two puckered layers (ABAB…) in HD and three (ABCABC…) in case of CD. The latter has slightly lower energy, essentially due to favourable long-range van der Waals interaction along the hexagonal direction. This means HD is a metastable structure that may occur if favoured by the kinetics of the shock experiments at certain intermediate pressures and high temperatures.

Previously the generalized solid-state nudged elastic band method has been used to study[49] HD and CD phase transformations, which indicates the transition pathways. Another study of the graphite-to-diamond transition is performed[29] wherein the transformation occurs through nucleation. Earlier AIMD simulations indicated[31] that transformation of HG to HD/CD is through sliding of graphite planes into an unusual orthorhombic stacking, from which an abrupt collapse and puckering of the planes lead to the transformation. Our AIMD simulations show that once the volume is sufficiently reduced at high pressure, the atoms move cooperatively and swiftly in a time duration of less than a pico-second leading to the transformation from HG to HD/CD. The same mechanism might not apply at lower pressures that may involve nucleation and growth. However, we note that the transition occurs swiftly in shock experiments involving both high pressure and temperature.

Further, in order to understand whether there is any soft phonon mode corresponding to the high-pressure phase transformation, we have calculated (**Fig. 4(a,b)**) the pressure dependence of phonon modes in HG. We find two soft modes at the Brillouin zone center in HG at high pressure when the c-lattice parameter approaches those of the HD/CD phases. One soft mode is of the degenerate $E_g$ symmetry and involves (**Fig. 4(c)**) sliding of the two graphitic layers in opposite directions in the a-b plane, while the other mode of $B_{1g}$ symmetry involves (**Fig. 4(c)**) displacement of the carbon atoms along the c-axis producing puckering of the layers. It seems coupling of both these soft modes brings about the phase transition mechanism of HG to HD/CD. Our AIMD simulations discussed above also show similar mechanism of the phase transitions. It is often said that the sliding of layers in HG is rather easy due to the weak van der Waals forces between them. However, we may note that the sliding phonon mode in HG is not a soft mode until the layer spacing is compressed to be close to that in HD/CD. At such a small layer spacing, the carbon atoms find it easy to switch from $sp^2$ to $sp^3$ bonding.

In summary, ab-initio molecular dynamics calculations have provided the mechanism of phase transition of hexagonal graphite to hexagonal or cubic diamond. We show that during the phase transition there is large cooperative displacement of atoms in very short time of about 0.2 ps. The transition paths of both HG to HD and HG to CD involve sliding of the graphite layers by about 0.7 Å with simultaneous puckering of the layers by about ±0.25 Å. The pattern of the sliding of the layers is different in case of the transition to HD and CD. The HD transition is achieved by sliding alternate layers in opposite directions along [2, 1, 0], while the CD transition involves sliding of pairs of layers along the [1, 0, 0] or equivalent directions. We find soft phonon modes occur at the Brillouin zone center in HG at high pressure when the c-lattice parameter approaches that of the HD/CD phases. These modes involve the sliding and puckering of the layers similar to that found in the AIMD simulations. We have also calculated the Gibbs free energies, which reveal important contributions of phonons, especially the vibrational entropy in stabilizing the HG phase up to higher pressures than the very low values predicted from previous enthalpy calculations, thus removing an important discrepancy between theory and experiments. The CD has slightly lower free energy than the HD, essentially due to the different layer sequence along the hexagonal-axis and favourable long-range van der Waals attraction for CD. However, phase transitions from HG to both HD and CD have been reported in experiments.

## ACKNOWLEDGEMENT

The use of ANUPAM super-computing facility at BARC is acknowledged. SLC acknowledges the financial support of the Indian National Science Academy for the INSA Senior Scientist position.

FIG. 1 (Color online) (a,b) The phonon density of states of hexagonal graphite (HG), hexagonal diamond (HG) and cubic diamond (CD): ab-initio calculations (lines), and experimental data (symbols) for HG [37] and CD [39]. The calculated spectra are broadened by a Gaussian of full-width-at-half-maximum of 6 meV. (c) The calculated volume ($V/V_o$) as a function of pressure at 0 K, normalized to the volume ($V_o$) of HG at 0 GPa. For comparison, the experimental data are shown for HG (solid magenta circles [44])) and HD (open red circle[14], solid red circles[16]) (d) The calculated Gibbs free energy (G) for 4 atoms as function of pressure in HG, HD and CD at 2000 K.

FIG. 2 (Color online) Transformation of hexagonal graphite (HG) to hexagonal diamond (HD) at 5500 K, and HG to cubic diamond (CD) at 300 K. The AIMD simulations are performed with NVT ensemble at constant lattice parameters corresponding to that of HD/CD at 49 GPa. The time dependence of the fractional coordinates $(I(x,y,z))$ of the atoms in the hexagonal unit cells, starting with the coordinates in HG. The x, y and z are the fractional coordinates along the a-, b- and c-axis, respectively, with reference to the HG unit cell. Dashed lines correspond to the fractional coordinates in the transformed phase. For the transitions from HG to HD and CD, the coordinates of the 4 and 6 atoms are plotted respectively, which form the hexagonal unit cell of the transformed phase. These coordinates are obtained from average over equivalent atoms in supercells of HG of size 4×4×1 and 3×3×3 respectively.

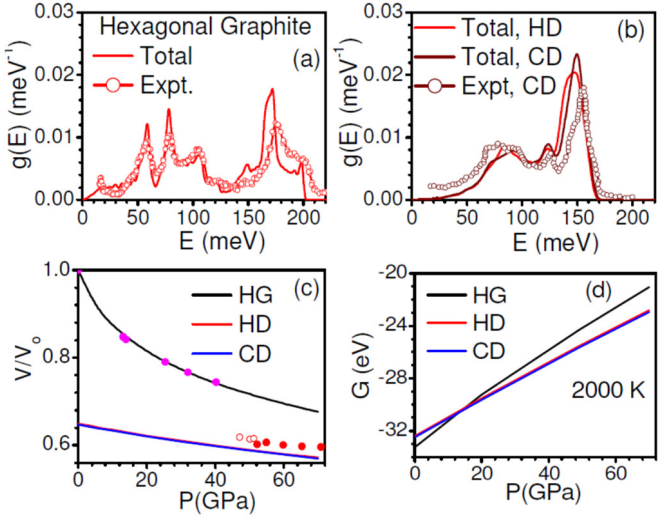

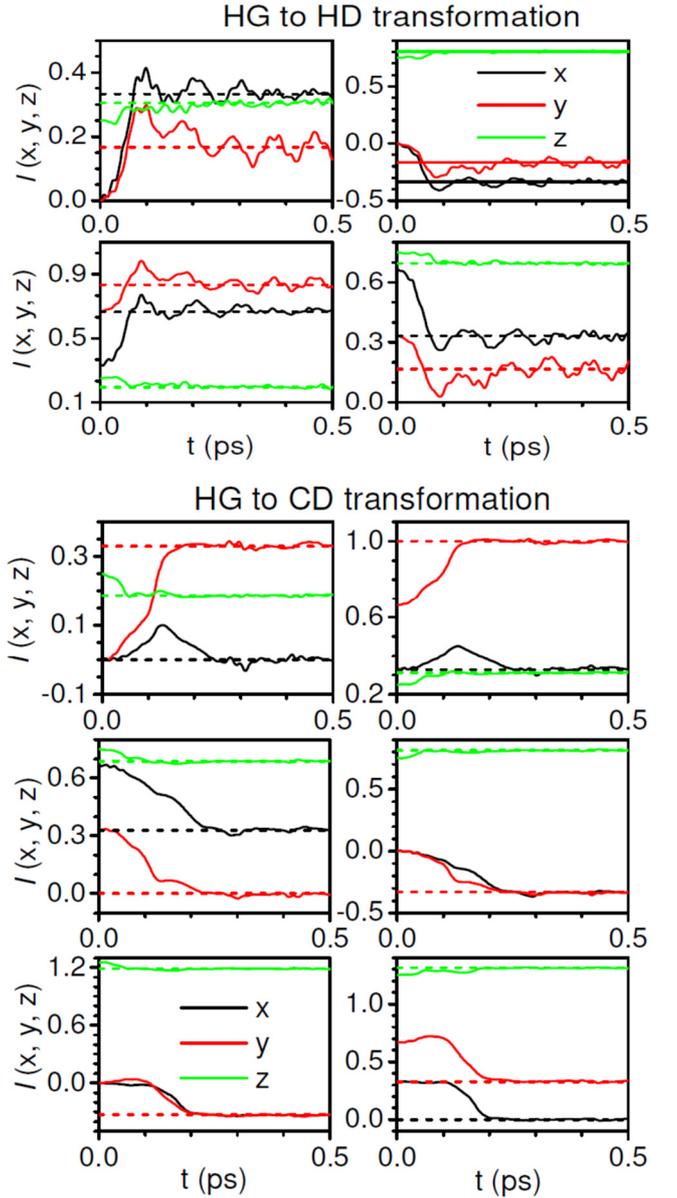

FIG. 3 (Color online) The snapshots of atoms during transformation of hexagonal graphite to hexagonal diamond (HD) (left panel) and cubic diamond (CD) (right panel). The time after the start of the simulation is indicated below each snapshot.

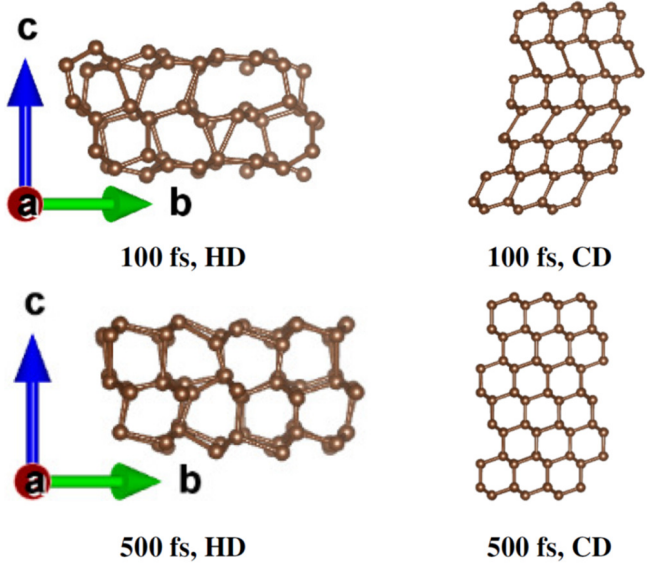

FIG. 4 (Color online) (a) Pressure dependence of low energy phonon modes in hexagonal graphite (HG) along (001); black line for a= 2.285 Å, c= 4.213 Å; blue line for a= 2.285, c= 4.183; and red line for a= 2.280 Å, c= 4.183 Å. (b) The energy of the two zone-centre soft phonon modes is calculated as function of pressure, and plotted as a function of the lattice parameters; black line- $E_g$ mode, red line- $B_{1g}$ mode. (c) The atomic displacement patterns of the two phonon modes. (Left) the mode of degenerate $E_g$ symmetry involves sliding of HG layers; (Right) the mode of $B_{1g}$ symmetry involves puckering of the layers.

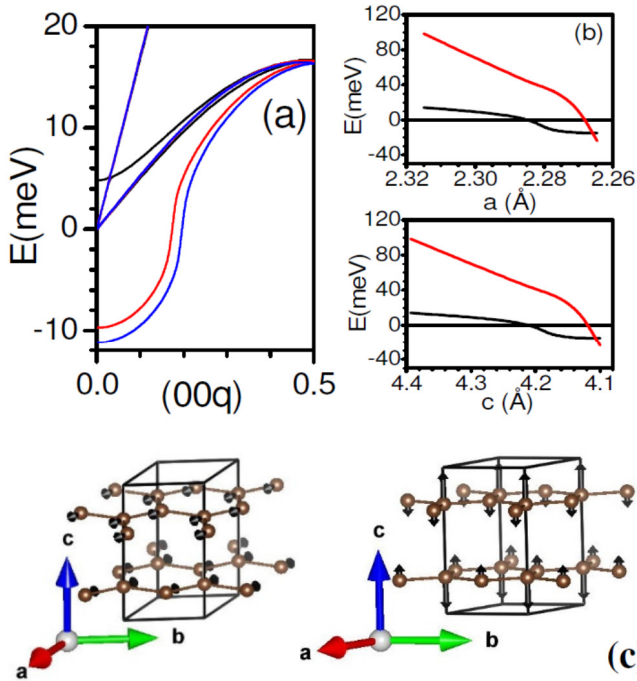